\documentclass[12pt]{iopart}
\usepackage{citesort}
\usepackage{epsfig}

\newcommand{\gtrsim}{\,\rlap{\lower3.5pt\hbox{$\mathchar\sim$}}
\raise1pt\hbox{$>$}\,}
\newcommand{\lesssim}{\,\rlap{\lower3.5pt\hbox{$\mathchar\sim$}}
\raise1pt\hbox{$<$}\,}

\begin{document}

\title{Neutrinos in Cosmology}
\author{Steen Hannestad}
\address{Department of Physics, University of Southern Denmark\\
Campusvej 55, DK-5230 Odense M, Denmark}

\date{{\today}}

\begin{abstract}
The current status of neutrino cosmology is reviewed, from the
question of neutrino decoupling and the presence of sterile
neutrinos to the effects of neutrinos on the cosmic microwave
background and large scale structure. Particular emphasis is put
on cosmological neutrino mass measurements.
\end{abstract}
\maketitle

%%%%%%%%%%%%%%%%%%%%%%%%%%%%%%%%%%%%%%%%%%%%%%%%%%%%%%%%%%%%%%%%%%%%%%
\section{Introduction} %%%%%%%%%%%%%%%%%%%%%%%%%%%%%%%%%%%%%%%%%%%%%%%
%%%%%%%%%%%%%%%%%%%%%%%%%%%%%%%%%%%%%%%%%%%%%%%%%%%%%%%%%%%%%%%%%%%%%%

Next to photons neutrinos are the most abundant particles in the
universe. This means they have a profound impact on many different
aspects of cosmology, from the question of leptogenesis in the
very early universe, over big bang nucleosynthesis, to late time
structure formation. In the present review I focus mainly on
late-time aspects of neutrino cosmology, and particularly on
issues relevant to cosmological bounds on the neutrino mass.

The absolute value of neutrino masses are very difficult to
measure experimentally. On the other hand, mass differences
between neutrino mass eigenstates, $(m_1,m_2,m_3)$, can be
measured in neutrino oscillation experiments.

The combination of all currently available data suggests two
important mass differences in the neutrino mass hierarchy. The
solar mass difference of $\delta m_{12}^2 \simeq 7 \times 10^{-5}$
eV$^2$ and the atmospheric mass difference $\delta m_{23}^2 \simeq
2.6 \times 10^{-3}$ eV$^2$
\cite{Maltoni:2003da,Aliani:2003ns,deHolanda:2003nj}.

In the simplest case where neutrino masses are hierarchical these
results suggest that $m_1 \sim 0$, $m_2 \sim \delta m_{\rm
solar}$, and $m_3 \sim \delta m_{\rm atmospheric}$. If the
hierarchy is inverted
\cite{Kostelecky:1993dm,Fuller:1995tz,Caldwell:1995vi,Bilenky:1996cb,King:2000ce,He:2002rv}
one instead finds $m_3 \sim 0$, $m_2 \sim \delta m_{\rm
atmospheric}$, and $m_1 \sim \delta m_{\rm atmospheric}$. However,
it is also possible that neutrino masses are degenerate
\cite{Ioannisian:1994nx,Bamert:vc,Mohapatra:1994bg,Minakata:1996vs,%
Vissani:1997pa,Minakata:1997ja,Ellis:1999my,Casas:1999tp,Casas:1999ac,%
Ma:1999xq,Adhikari:2000as}, $m_1 \sim m_2 \sim m_3 \gg \delta
m_{\rm atmospheric}$, in which case oscillation experiments are
not useful for determining the absolute mass scale.

Experiments which rely on kinematical effects of the neutrino mass
offer the strongest probe of this overall mass scale. Tritium
decay measurements have been able to put an upper limit on the
electron neutrino mass of 2.3 eV (95\% conf.) \cite{kraus}.
However, cosmology at present yields an much stronger limit which
is also based on the kinematics of neutrino mass.

Very interestingly there is also a claim of direct detection of
neutrinoless double beta decay in the Heidelberg-Moscow experiment
\cite{Klapdor-Kleingrothaus:2001ke,Klapdor-Kleingrothaus:2004wj},
corresponding to an effective neutrino mass in the $0.1-0.9$ eV
range. If this result is confirmed then it shows that neutrino
masses are almost degenerate and well within reach of cosmological
detection in the near future.

Another important question which can be answered by cosmological
observations is how large the total neutrino energy density is.
Apart from the standard model prediction of three light neutrinos,
such energy density can be either in the form of additional,
sterile neutrino degrees of freedom, or a non-zero neutrino
chemical potential.

The paper is divided into sections in the following way: In
section 2 I review the present cosmological data which can be used
for analysis of neutrino physics. In section 3 I discuss neutrino
physics around the epoch of neutrino decoupling at a temperature
of roughly 1 MeV, including the relation between neutrinos and Big
Bang nucleosynthesis. Section 4 discusses neutrinos as dark matter
particles, including mass constraints on light neutrinos, and
sterile neutrino dark matter. Section 5 contains a relatively
short review of neutrino physics in the very early universe from
the perspective of leptogenesis. Finally, section 6 contains a
discussion.

%%%%%%%%%%%%%%%%%%%%%%%%%%%%%%%%%%%%%%%%%%%%%%%%%%%%%%%%%%%%%%%%%%%%%%
\section{Cosmological data} %%%%%%%%%%%%%%%%%%%%%%%%%%%%%%%%%%%%%%%%%%
%%%%%%%%%%%%%%%%%%%%%%%%%%%%%%%%%%%%%%%%%%%%%%%%%%%%%%%%%%%%%%%%%%%%%%

\paragraph{Large Scale Structure (LSS) --}

At present there are two large galaxy surveys of comparable size,
the Sloan Digital Sky Survey (SDSS)
\cite{Tegmark:2003uf,Tegmark:2003ud} and the 2dFGRS (2~degree
Field Galaxy Redshift Survey) \cite{2dFGRS}. Once the SDSS is
completed in 2005 it will be significantly larger and more
accurate than the 2dFGRS. At present the two surveys are, however,
comparable in precision.

\paragraph{Cosmic Microwave Background (CMB) --}

The temperature fluctuations are conveniently described in terms
of the spherical harmonics power spectrum $C_{T,l} \equiv \langle
|a_{lm}|^2 \rangle$, where $\frac{\Delta T}{T} (\theta,\phi) =
\sum_{lm} a_{lm}Y_{lm}(\theta,\phi)$.  Since Thomson scattering
polarizes light, there are also power spectra coming from the
polarization. The polarization can be divided into a curl-free
$(E)$ and a curl $(B)$ component, yielding four independent power
spectra: $C_{T,l}$, $C_{E,l}$, $C_{B,l}$, and the $T$-$E$
cross-correlation $C_{TE,l}$.

The WMAP experiment has reported data only on $C_{T,l}$ and
$C_{TE,l}$ as described in
Refs.~\cite{Spergel:2003cb,Bennett:2003bz,Kogut:2003et,%
Hinshaw:2003ex,Verde:2003ey,Peiris:2003ff}. Foreground
contamination has already been subtracted from their published
data.

In addition to the WMAP experiment there are a number of other
current CMB experiments, both ground and balloon based. Wang {\it
et al.} \cite{wang3} have provided a compilation of various data
sets, and in addition to these there is the ACBAR experiment
\cite{Kuo:2002ua} which has measured the CMB at small scales.

\paragraph{Other data --}

Apart from CMB and LSS data there are a number of other
cosmological measurements of importance to neutrino cosmology. One
is the measurement of the Hubble constant by the HST Hubble Key
Project, $H_0=72 \pm 8~{\rm km}~{\rm s}^{-1}~{\rm Mpc}^{-1}$
\cite{Freedman:2000cf}.

The constraint on the matter density coming from measurements of
distant type Ia supernovae is also important for neutrino physics.
The most recent result is from the Supernova Cosmology Project
\cite{Knop:2003iy} and yields $\Omega_m = 0.25^{+0.07}_{-0.06}$
(statistical) $\pm0.04$ (identified systematics).

%%%%%%%%%%%%%%%%%%%%%%%%%%%%%%%%%%%%%%%%%%%%%%%%%%%%%%%%%%%%%%%%%%%%%%
\section{Neutrino Decoupling} %%%%%%%%%%%%%%%%%%%%%%%%%%%%%%%%%%%%%%%%
%%%%%%%%%%%%%%%%%%%%%%%%%%%%%%%%%%%%%%%%%%%%%%%%%%%%%%%%%%%%%%%%%%%%%%

\subsection{Standard model}

In the standard model neutrinos interact via weak interactions
with $e^+$ and $e^-$. In the absence of oscillations neutrino
decoupling can be followed via the Boltzmann equation for the
single particle distribution function \cite{kolb}
\begin{equation}
\frac{\partial f}{\partial t} - H p \frac{\partial f}{\partial p}
= C_{\rm coll}, \label{eq:boltz}
\end{equation}
where $C_{\rm coll}$ represents all elastic and inelastic
interactions. In the standard model all these interactions are $2
\leftrightarrow 2$ interactions in which case the collision
integral for process $i$ can be written
\begin{eqnarray}
C_{\rm coll,i} (f_1) & = & \frac{1}{2E_1} \int \frac{d^3 {\bf
p}_2}{2E_2 (2\pi)^3} \frac{d^3 {\bf p}_3}{2E_3 (2\pi)^3} \frac{d^3
{\bf p}_4}{2E_4 (2\pi)^3} \nonumber \\
&& \,\, \times (2\pi)^4 \delta^4
(p_1+p_2-p_3+p_4)\Lambda(f_1,f_2,f_3,f_4) S |M|^2_{12 \to 34,i},
\end{eqnarray}
where $S |M|^2_{12 \to 34,i}$ is the spin-summed and averaged
matrix element including the symmetry factor $S=1/2$ if there are
identical particles in initial or final states. The phase-space
factor is $\Lambda(f_1,f_2,f_3,f_4) = f_3 f_4 (1-f_1)(1-f_2) - f_1
f_2 (1-f_3)(1-f_4)$.

The matrix elements for all relevant processes can for instance be
found in Ref.~\cite{Hannestad:1995rs}. If Maxwell-Boltzmann
statistics is used for all particles, and neutrinos are assumed to
be in complete scattering equilbrium so that they can be
represented by a single temperature, then the collision integral
can be integrated to yield the average annihilation rate for a
neutrino
\begin{equation}
\Gamma = \frac{16 G_F^2}{\pi^3} (g_L^2 + g_R^2) T^5,
\end{equation}
where
\begin{equation}
g_L^2 + g_R^2 = \cases{\sin^4 \theta_W + (\frac{1}{2}+\sin^2
\theta_W)^2 & for $\nu_e$ \cr \sin^4 \theta_W +
(-\frac{1}{2}+\sin^2 \theta_W)^2 & for $\nu_{\mu,\tau}$}.
\end{equation}

This rate can then be compared with the Hubble expansion rate
\begin{equation}
H = 1.66 g_*^{1/2} \frac{T^2}{M_{\rm Pl}}
\end{equation}

 to
find the decoupling temperature from the criterion $\left. H =
\Gamma \right|_{T=T_D}$. From this one finds that $T_D(\nu_e)
\simeq 2.4$ MeV, $T_D(\nu_{\mu,\tau}) \simeq 3.7$ MeV, when $g_*
=10.75$, as is the case in the standard model.

This means that neutrinos decouple at a temperature which is
significantly higher than the electron mass. When $e^+e^-$
annihilation occurs around $T \sim m_e/3$, the neutrino
temperature is unaffected whereas the photon temperature is heated
by a factor $(11/4)^{1/3}$. The relation $T_\nu/T_\gamma =
(4/11)^{1/3} \simeq 0.71$ holds to a precision of roughly one
percent. The main correction comes from a slight heating of
neutrinos by $e^+e^-$ annihilation, as well as finite temperature
QED effects on the photon propagator
\cite{Dicus:1982bz,Rana:1991xk,herrera,Dolgov:1992qg,Dodelson:1992km,%
Fields:1993zb,Hannestad:1995rs,Dolgov:1997mb,Dolgov:1999sf,gnedin,%
Esposito:2000hi,Steigman:2001px,Mangano:2001iu,osc}.

\subsection{Big Bang nucleosynthesis and the number of neutrino species}

Shortly after neutrino decoupling the weak interactions which keep
neutrons and protons in statistical equilibrium freeze out. Again
the criterion $\left. H = \Gamma \right|_{T=T_{\rm freeze}}$ can
be applied to find that $T_{\rm freeze} \simeq 0.5 g_*^{1/6}$ MeV
\cite{kolb}.

Eventually, at a temperature of roughly 0.2 MeV deuterium starts
to form, and very quickly all free neutrons are processed into
$^4$He. The final helium abundance is therefore roughly given by
\begin{equation}
Y_P \simeq \left. \frac{2 n_n/n_p}{1+n_n/n_p} \right|_{T\simeq 0.2
\,\, {\rm MeV}}.
\end{equation}

$n_n/n_p$ is determined by its value at freeze out, roughly by the
condition that $n_n/n_p|_{T=T_{\rm freeze}} \sim
e^{-(m_n-m_p)/T_{\rm freeze}}$.

Since the freeze-out temperature is determined by $g_*$ this in
turn means that $g_*$ can be inferred from a measurement of the
helium abundance. However, since $Y_P$ is a function of both
$\Omega_b h^2$ and $g_*$ it is necessary to use other measurements
to constrain $\Omega_b h^2$ in order to find a bound on $g_*$. One
customary method for doing this has been to use measurements of
primordial deuterium to infer $\Omega_b h^2$ and from that
calculate a bound on $g_*$. Usually such bounds are expressed in
terms of the equivalent number of neutrino species, $N_\nu \equiv
\rho/\rho_{\nu_0}$, instead of $g_*$. The exact value of the bound
is quite uncertain because there are different and inconsistent
measurements of the primordial helium abundance (see for instance
Ref.~\cite{Barger:2003zg} for a discussion of this issue). The
most recent analyses are \cite{Barger:2003zg} where a value of
$1.7 \leq N_\nu \leq 3.0$ (95\% C.L.) was found and
\cite{Cuoco:2003cu} which found the result $N_\nu =
2.5^{+1.1}_{-0.9}$. The difference in these results can be
attributed to different assumptions about uncertainties in the
primordial helium abundance.

Another interesting parameter which can be constrained by the same
argument is the neutrino chemical potential, $\xi_\nu=\mu_\nu/T$
\cite{Kang:xa,Kohri:1996ke,Orito:2002hf,Ichikawa:2002vn}. At first
sight this looks like it is completely equivalent to constraining
$N_\nu$. However, this is not true because a chemical potential
for electron neutrinos directly influences the $n-p$ conversion
rate. Therefore the bound on $\xi_{\nu_e}$ from BBN alone is
relatively stringent ($-0.1 \lesssim \xi_{\nu_e} \lesssim 1$
\cite{Kang:xa}) compared to that for muon and tau neutrinos
($\left|\xi_{\nu_{\mu,\tau}}\right| \lesssim 7$ \cite{Kang:xa}).
However, as will be seen in the next section, neutrino
oscillations have the effect of almost equilibrating the neutrino
chemical potentials prior to BBN, completely changing this
conclusion.

\subsection{The number of neutrino species - joint CMB and BBN analysis}

The BBN bound on the number of neutrino species presented in the
previous section can be complemented by a similar bound from
observations of the CMB and large scale structure. The CMB depends
on $N_\nu$ mainly because of the early Integrated Sachs Wolfe
effect which increases fluctuation power at scales slightly larger
than the first acoustic peak. The large scale structure spectrum
depends on $N_\nu$ because the scale of matter-radiation equality
is changed by varying $N_\nu$.

Several recent papers have analyzed WMAP and 2dF data for bounds
on $N_\nu$
\cite{Crotty:2003th,Hannestad:2003xv,Pierpaoli:2003kw,Barger:2003zg,%
Cuoco:2003cu}, and some of the bounds are listed in Table
\ref{table:nnu}. Recent analyses combining BBN, CMB, and large
scale structure data can be found in
\cite{Hannestad:2003xv,Barger:2003zg}, and these results are also
listed in Table \ref{table:nnu}.

Common for all the bounds is that $N_\nu=0$ is ruled out by both
BBN and CMB/LSS. This has the important consequence that the
cosmological neutrino background has been positively detected, not
only during the BBN epoch, but also much later, during structure
formation.

\begin{table}
\begin{center}
\caption{Various recent limits on the effective number of neutrino
species, as well as the data used.}
\begin{tabular}{@{}lll}
\hline

Ref. & Bound on $N_\nu$ & Data used \\
\hline

Crotty et al. \cite{Crotty:2003th} & $1.4 \leq N_\nu \leq 6.8$ &
CMB, LSS \\

Hannestad \cite{Hannestad:2003xv} & $0.9 \leq N_\nu \leq 7.0$ &
CMB, LSS \\

Pierpaoli \cite{Pierpaoli:2003kw} & $1.9 \leq N_\nu \leq 6.62$ &
CMB, LSS \\

Barger et al. \cite{Barger:2003zg} & $0.9 \leq N_\nu \leq 8.3$ &
CMB \\

Hannestad \cite{Hannestad:2003xv} & $2.3 \leq N_\nu \leq 3.0$ &
CMB, LSS, BBN \\

Barger et al. \cite{Barger:2003zg} & $1.7 \leq N_\nu \leq 3.0$ &
CMB, BBN \\

\hline
\end{tabular}
\end{center}
\label{table:nnu}
\end{table}

\subsection{The effect of oscillations}

In the previous section the one-particle distribution function,
$f$, was used to describe neutrino evolution. However, for
neutrinos the mass eigenstates are not equivalent to the flavour
eigenstates because neutrinos are mixed. Therefore the evolution
of the neutrino ensemble is not in general described by the three
scalar functions, $f_i$, but rather by the evolution of the
neutrino density matrix, $\rho \equiv \psi \psi^\dagger$, the
diagonal elements of which correspond to $f_i$.

For three-neutrino oscillations the formalism is quite
complicated. However, the difference in $\Delta m_{12}$ and
$\Delta m_{23}$, as well as the fact that $\sin 2 \theta_{13} \ll
1$ means that the problem effectively reduces to a $2 \times 2$
oscillation problem in the standard model. A detailed account of
the physics of neutrino oscillations in the early universe is
outside the scope of the present paper, however an excellent and
very thorough review can be found in Ref.~\cite{dolgov}

Without oscillations it is possible to compensate a very large
chemical potential for muon and/or tau neutrinos with a small,
negative electron neutrino chemical potential \cite{Kang:xa}.
However, since neutrinos are almost maximally mixed a chemical
potential in one flavour can be shared with other flavours, and
the end result is that during BBN all three flavours have almost
equal chemical potential. This in turn means that the bound on
$\nu_e$ applies to all species so that
\cite{lunardini,Pastor:2001iu,Dolgov:2002ab,Abazajian:2002qx,Wong:2002fa}.

\begin{equation}
|\xi_i| = \frac{|\eta_i|}{T} \lesssim 0.15
\end{equation}
for $i=e,\mu,\tau$.

In models where sterile neutrinos are present even more remarkable
oscillation phenomena can occur. However, I do not discuss this
possibility further, except for the possibility of sterile
neutrino warm dark matter, and instead refer to the review
\cite{dolgov}.

\subsection{Low reheating temperature and neutrinos}

In most models of inflation the universe enters the normal,
radiation dominated epoch at a reheating temperature, $T_{\rm
RH}$, which is of order the electroweak scale or higher. However,
in principle it is possible that this reheating temperature is
much lower, of order MeV. This possibility has been studied many
times in the literature, and a very general bound of $T_{\rm RH}
\gtrsim 1$ MeV has been found
\cite{Kawasaki:1999na,Kawasaki:2000en,Giudice:2000ex,Giudice:2000dp}

This very conservative bound comes from the fact that the light
element abundances produced by big bang nucleosynthesis disagree
with observations if the universe if matter dominated during BBN.
However, a somewhat more stringent bound can be obtained by
looking at neutrino thermalization during reheating. If a scalar
particle is responsible for reheating then direct decay to
neutrinos is suppressed because of the necessary helicity flip.
This means that if the reheating temperature is too low neutrinos
never thermalize. If this is the case then BBN predicts the wrong
light element abundances. However, even if the heavy particle has
a significant branching ratio into neutrinos there are problems
with BBN. The reason is that neutrinos produced in decays are born
with energies which are much higher than thermal. If the reheating
temperature is too low then a population of high energy neutrinos
will remain and also lead to conflict with observed light element
abundances. A recent analysis showed that in general the reheating
temperature cannot be below roughly 4 MeV \cite{Hannestad:2004px}.

%%%%%%%%%%%%%%%%%%%%%%%%%%%%%%%%%%%%%%%%%%%%%%%%%%%%%%%%%%%%%%%%%%%%%%
\section{Neutrino Dark Matter} %%%%%%%%%%%%%%%%%%%%%%%%%%%%%%%%%%%%%%%
%%%%%%%%%%%%%%%%%%%%%%%%%%%%%%%%%%%%%%%%%%%%%%%%%%%%%%%%%%%%%%%%%%%%%%

Neutrinos are a source of dark matter in the present day universe
simply because they contribute to $\Omega_m$. The present
temperature of massless standard model neutrinos is $T_{\nu,0} =
1.95 \, K = 1.7 \times 10^{-4}$ eV, and any neutrino with $m \gg
T_{\nu,0}$ behaves like a standard non-relativistic dark matter
particle.

The present contribution to the matter density of $N_\nu$ neutrino
species with standard weak interactions is given by
\begin{equation}
\Omega_\nu h^2 = N_\nu \frac{m_\nu}{92.5 \, {\rm eV}}
\end{equation}
Just from demanding that $\Omega_\nu \leq 1$ one finds the bound
\cite{Gershtein:gg,Cowsik:gh}
\begin{equation}
m_\nu \lesssim \frac{46 \, {\rm eV}}{N_\nu} \label{eq:mnu}
\end{equation}

\subsection{The Tremaine-Gunn bound}

If neutrinos are the main source of dark matter, then they must
also make up most of the galactic dark matter. However, neutrinos
can only cluster in galaxies via energy loss due to gravitational
relaxation since they do not suffer inelastic collisions. In
distribution function language this corresponds to phase mixing of
the distribution function \cite{Tremaine:we}. By using the theorem
that the phase-mixed or coarse grained distribution function must
explicitly take values smaller than the maximum of the original
distribution function one arrives at the condition
\begin{equation}
f_{\rm CG} \leq f_{\nu,{\rm max}} = \frac{1}{2}
\end{equation}
Because of this upper bound it is impossible to squeeze neutrino
dark matter beyond a certain limit \cite{Tremaine:we}. For the
Milky Way this means that the neutrino mass must be larger than
roughly 25 eV {\it if} neutrinos make up the dark matter. For
irregular dwarf galaxies this limit increases to 100-300 eV
\cite{Madsen:mz,salucci}, and means that standard model neutrinos
cannot make up a dominant fraction of the dark matter. This bound
is generally known as the Tremaine-Gunn bound.

Note that this phase space argument is a purely classical
argument, it is not related to the Pauli blocking principle for
fermions (although, by using the Pauli principle $f_\nu \leq 1$
one would arrive at a similar, but slightly weaker limit for
neutrinos). In fact the Tremaine-Gunn bound works even for bosons
if applied in a statistical sense \cite{Madsen:mz}, because even
though there is no upper bound on the fine grained distribution
function, only a very small number of particles reside at low
momenta (unless there is a condensate). Therefore, although the
exact value of the limit is model dependent, limit applies to any
species that was once in thermal equilibrium. A notable
counterexample is non-thermal axion dark matter which is produced
directly into a condensate.

\subsection{Neutrino hot dark matter}

A much stronger upper bound on the neutrino mass than the one in
Eq.~(\ref{eq:mnu}) can be derived by noticing that the thermal
history of neutrinos is very different from that of a WIMP because
the neutrino only becomes non-relativistic very late.

In an inhomogeneous universe the Boltzmann equation for a
collisionless species is \cite{MB}
\begin{equation}
L[f] = \frac{Df}{D\tau} = \frac{\partial f}{\partial \tau}
 + \frac{dx^i}{d\tau}\frac{\partial f}{\partial x^i} +
\frac{dq^i}{d\tau}\frac{\partial f}{\partial q^i} = 0,
\end{equation}
where $\tau$ is conformal time, $d \tau = dt/a$, and $q^i = a p^i$
is comoving momentum. The second term on the right-hand side has
to do with the velocity of the distribution in a given spatial
point and the third term is the cosmological momentum redshift.

Following Ma and Bertschinger \cite{MB} this can be rewritten as
an equation for $\Psi$, the perturbed part of $f$
\begin{equation}
f(x^i,q^i,\tau) = f_0(q) \left[ 1 + \Psi(x^i,q^i,\tau) \right]
\end{equation}

In synchronous gauge that equation is

\begin{equation}
\frac{1}{f_0}[f] = \frac{\partial \Psi}{\partial \tau} + i
\frac{q}{\epsilon} \mu \Psi + \frac{d \ln f_0}{d \ln q}
\left[\dot{\eta}-\frac{\dot{h}+6\dot{\eta}} {2} \mu^2 \right] =
\frac{1}{f_0} C[f],
\end{equation}
where $q^j = q n^j$, $\mu \equiv n^j \hat{k}_j$, and $\epsilon =
(q^2 + a^2 m^2)^{1/2}$. $k^j$ is the comoving wavevector. $h$ and
$\eta$ are the metric perturbations, defined from the perturbed
space-time metric in synchronous gauge \cite{MB}
\begin{equation}
ds^2 = a^2(\tau) [-d\tau^2 + (\delta_{ij} + h_{ij})dx^i dx^j],
\end{equation}
\begin{equation}
h_{ij} = \int d^3 k e^{i \vec{k}\cdot\vec{x}}\left(\hat{k}_i
\hat{k}_j h(\vec{k},\tau) +(\hat{k}_i \hat{k}_j - \frac{1}{3}
\delta_{ij}) 6 \eta (\vec{k},\tau) \right).
\end{equation}

Expanding this in Legendre polynomials one arrives at a set of
hierarchy equations
\begin{eqnarray}
\dot{\delta} & = & -\frac{4}{3} \theta - \frac{2}{3} \dot h \nonumber \\
\dot{\theta} & = & k^2\left(\frac{\delta}{4} - \sigma \right) \nonumber \\
2 \dot{\sigma} & = & \frac{8}{15} \theta - \frac{3}{15} k F_3
+ \frac{4}{15} \dot h + \frac{8}{5} \dot{\eta} \nonumber \\
\dot{F}_l & = & \frac{k}{2l+1} \left(l F_{l-1} - (l+1) F_{l+1}
\right)
\end{eqnarray}
For subhorizon scales ($\dot h = \dot \eta = 0$) this reduces to
the form
\begin{eqnarray}
\dot{\delta} & = & -\frac{4}{3} \theta \nonumber \\
\dot{\theta} & = & k^2\left(\frac{\delta}{4} - \sigma \right) \nonumber \\
2 \dot{\sigma} & = & \frac{8}{15} \theta - \frac{3}{15} k F_3 \nonumber \\
\dot{F}_l & = & \frac{k}{2l+1} \left(l F_{l-1} - (l+1) F_{l+1}
\right)
\end{eqnarray}

One should notice the similarity between this set of equations and
the evolution hierarchy for spherical Bessel functions. Indeed the
exact solution to the hierarchy is
\begin{equation}
F_{l}(k \tau) \sim j_l(k \tau)
\end{equation}
This shows that the solution for $\delta$ is an exponentially
damped oscillation. On small scales, $k > \tau$, perturbations are
erased.

This in intuitively understandable in terms of free-streaming.
Indeed the Bessel function solution comes from the fact that
neutrinos are considered massless. In the limit of CDM the
evolution hierarchy is truncated by the fact that $\theta=0$, so
that the CDM perturbation equation is simply $\dot \delta = -\dot
h/2$. For massless particles the free-streaming length is $\lambda
= c \tau$ which is reflected in the solution to the Boltzmann
hierarchy. Of course the solution only applies when neutrinos are
strictly massless. Once $T \sim m$ there is a smooth transition to
the CDM solution. Therefore the final solution can be separated
into two parts: 1) $k > \tau(T=m)$: Neutrino perturbations are
exponentially damped 2) $k < \tau(T=m)$: Neutrino perturbations
follow the CDM perturbations. Calculating the free streaming
wavenumber in a flat CDM cosmology leads to the simple numerical
relation (applicable only for $T_{\rm eq} \gg m \gg T_0$)
\cite{kolb}

\begin{equation}
\lambda_{\rm FS} \sim \frac{20~{\rm Mpc}}{\Omega_x h^2}
\left(\frac{T_x}{T_\nu}\right)^4 \left[1+\log \left(3.9
\frac{\Omega_x h^2}{\Omega_m h^2} \left(\frac{T_\nu}{T_x}\right)^2
\right)\right]\,. \label{eq:freestream}
\end{equation}

\begin{figure}[htbp]
\begin{center}
\epsfig{file=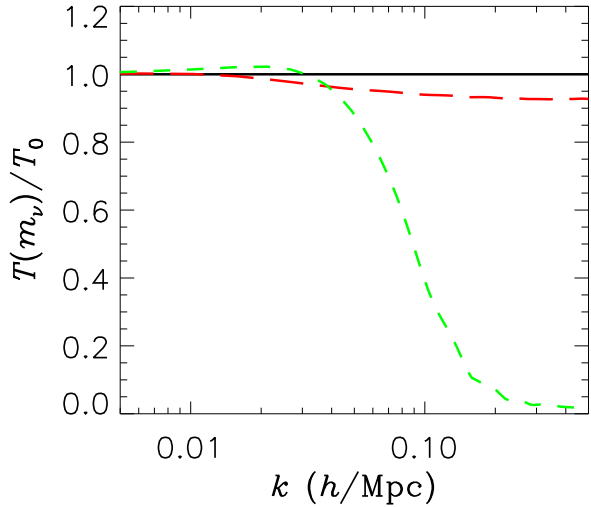,width=0.6\textwidth}
\end{center}
\bigskip
\caption{\label{fig:nutrans} The transfer function $T(k,t=t_0)$
for various different neutrino masses. The solid (black) line is
for $m_\nu=0$, the long-dashed for $m_\nu = 0.3$ eV, and the
dashed for $m_\nu=1$ eV.}
\end{figure}

In Fig.~\ref{fig:nutrans} I have plotted transfer functions for
various different neutrino masses in a flat $\Lambda$CDM universe
$(\Omega_m+\Omega_\nu+\Omega_\Lambda=1)$. The parameters used were
$\Omega_b = 0.04$, $\Omega_{\rm CDM} = 0.26 - \Omega_\nu$,
$\Omega_\Lambda = 0.7$, $h = 0.7$, and $n=1$.

When measuring fluctuations it is customary to use the power
spectrum, $P(k,\tau)$, defined as
\begin{equation}
P(k,\tau) = |\delta|^2(\tau).
\end{equation}
The power spectrum can be decomposed into a primordial part,
$P_0(k)$, and a transfer function $T(k,\tau)$,
\begin{equation}
P(k,\tau) = P_0(k) T(k,\tau).
\end{equation}
The transfer function at a particular time is found by solving the
Boltzmann equation for $\delta(\tau)$.

At scales much smaller than the free-streaming scale the present
matter power spectrum is suppressed roughly by the factor
\cite{Hu:1997mj}
\begin{equation}
\frac{\Delta P(k)}{P(k)} = \frac{\Delta
T(k,\tau=\tau_0)}{T(k,\tau=\tau_0)}\simeq -8
\frac{\Omega_\nu}{\Omega_m},
\end{equation}
as long as $\Omega_\nu \ll \Omega_m$. The numerical factor 8 is
derived from a numerical solution of the Boltzmann equation, but
the general structure of the equation is simple to understand. At
scales smaller than the free-streaming scale the neutrino
perturbations are washed out completely, leaving only
perturbations in the non-relativistic matter (CDM and baryons).
Therefore the {\it relative} suppression of power is proportional
to the ratio of neutrino energy density to the overall matter
density. Clearly the above relation only applies when $\Omega_\nu
\ll \Omega_m$, when $\Omega_\nu$ becomes dominant the spectrum
suppression becomes exponential as in the pure hot dark matter
model. This effect is shown for different neutrino masses in
Fig.~\ref{fig:nutrans}.

\begin{figure}[htbp]
\begin{center}
\hspace*{-1cm}\epsfig{file=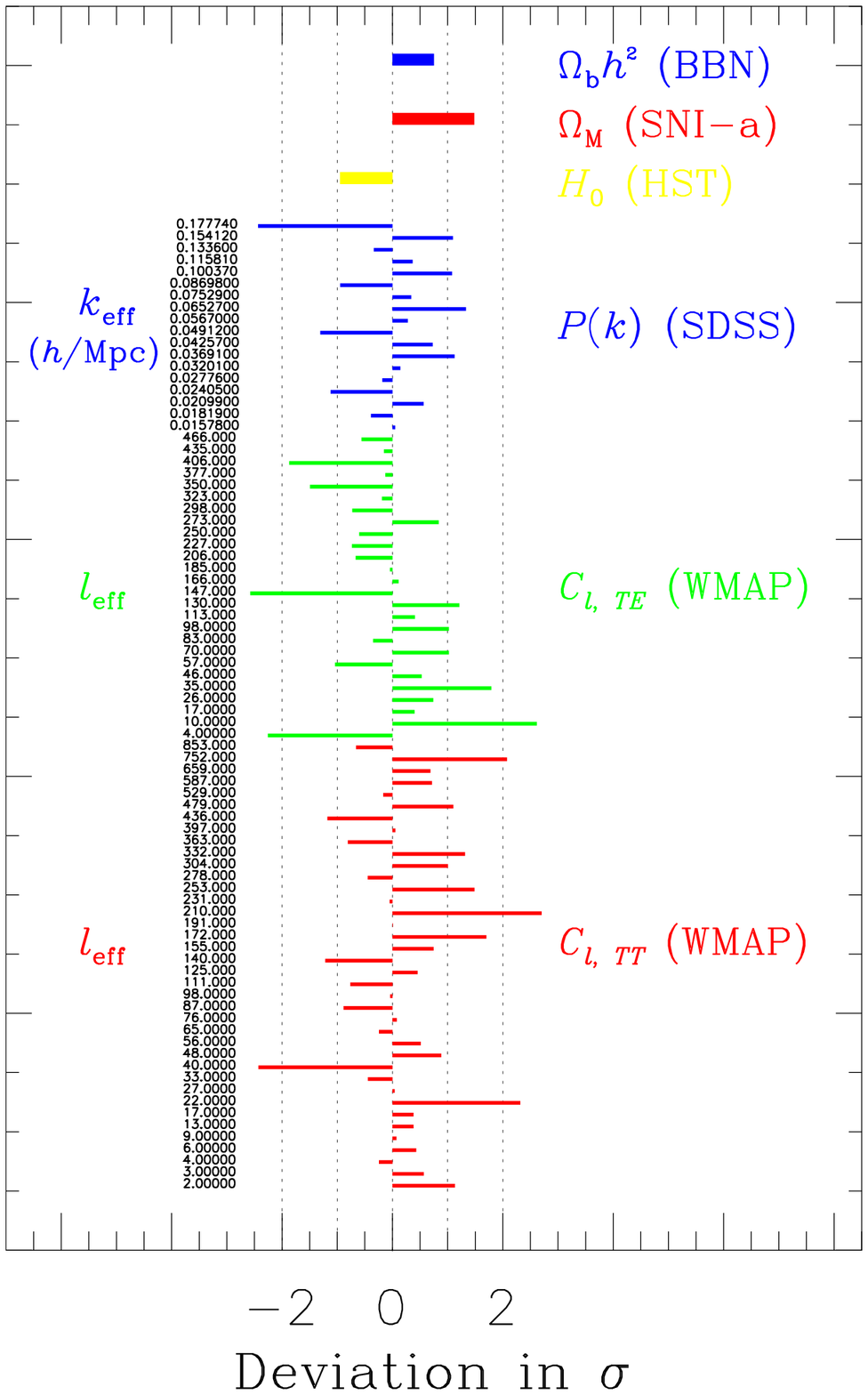,width=0.6\textwidth}\hspace*{-2cm}\epsfig{file=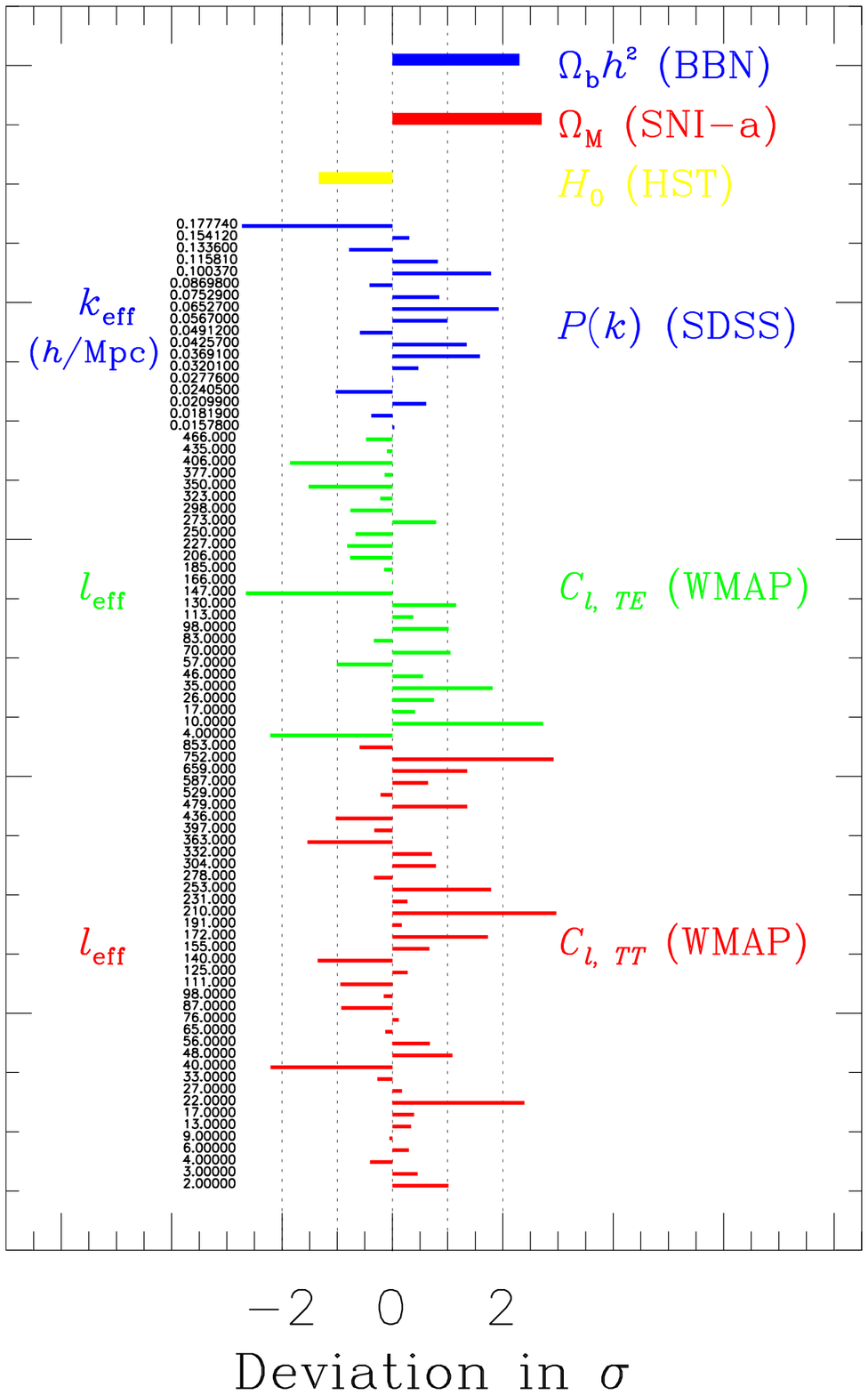,width=0.6\textwidth}
\end{center}
\bigskip
\caption{Deviation of the best fit models for $\sum m_\nu = 0$
(left) and $\sum m_\nu = 3$ eV (right). The CMB data shown are the
WMAP binned data (http://lambda.gsfc.nasa.gov). The deviation is
calculated as "Deviation in $\sigma$" $= (A_{\rm model} - A_{\rm
obs})/\sigma_{\rm obs}$.} \label{fig:pull}
\end{figure}

The effect of massive neutrinos on structure formation only
applies to the scales below the free-streaming length. For
neutrinos with masses of several eV the free-streaming scale is
smaller than the scales which can be probed using present CMB data
and therefore the power spectrum suppression can be seen only in
large scale structure data. On the other hand, neutrinos of sub-eV
mass behave almost like a relativistic neutrino species for CMB
considerations. The main effect of a small neutrino mass on the
CMB is that it leads to an enhanced early ISW effect. The reason
is that the ratio of radiation to matter at recombination becomes
larger because a sub-eV neutrino is still relativistic or
semi-relativistic at recombination. With the WMAP data alone it is
very difficult to constrain the neutrino mass, and to achieve a
constraint which is competitive with current experimental bounds
it is necessary to include LSS data from 2dF or SDSS. When this is
done the bound becomes very strong, somewhere in the range of 1 eV
for the sum of neutrino masses, depending on assumptions about
priors. In Table \ref{table:mass2} the present upper bound on the
neutrino mass from various analyses is quoted, as well as the
assumptions going into the derivation.

\begin{table}
\begin{center}
\caption{Various recent limits on the neutrino mass from cosmology
and the data sets used in deriving them. 1: WMAP data, 2: Other
CMB data, 3: 2dF data, 4: Constraint on $\sigma_8$ (different in
$4^a$ and $4^b$), 5: SDSS data, 6: Constraint on $H_0$.}
\begin{tabular}{@{}lll}
\hline

Ref. & Bound on $\sum m_\nu$ & Data used \\
\hline

Spergel et al. (WMAP) \cite{Spergel:2003cb}  &   0.69 eV     &   1,2,3,$4^a$,6 \\
Hannestad \cite{Hannestad:2003xv} &   1.01 eV     &   1,2,3,6 \\
Allen, Smith and Bridle \cite{Allen:2003pt} & $0.56^{+0.3}_{-0.26}$ eV & 1,2,3,$4^b$,6 \\
Tegmark et al. (SDSS) \cite{Tegmark:2003ud} & 1.8 eV & 1,5 \\
Barger et al. \cite{Barger:2003vs} & 0.75 eV & 1,2,3,5,6 \\
Crotty, Lesgourgues and Pastor \cite{Crotty:2004gm} & 1.0 (0.6) eV & 1,2,3,5 (6) \\

\hline
\end{tabular}
\end{center}
\label{table:mass2}
\end{table}

As can be gauged from this table, a fairly robust bound on the sum
of neutrino masses is at present somewhere around 1.0 eV,
depending somewhat on specific priors and data sets used.

It is also quite interesting to see what exactly provides this
bound. It is often stated that the neutrino mass bound comes from
large scale structure data, not from CMB because CMB probes larger
scales. However, LSS data alone provides no limit on $\sum m_\nu$
because of degeneracies with other parameters (this is discussed
in detail in Ref.~\cite{Elgaroy:2003yh}). On the other hand, WMAP
in itself also does not provide a strong limit on the neutrino
mass \cite{Tegmark:2003ud}, because neutrino mass only has a
limited effect on the scales probed by WMAP. Only the combination
of the two types of data allows for a determination of $\sum
m_\nu$ with any precision. Fig.~\ref{fig:pull} show deviation of
the best fit models for $\sum m_\nu = 0$ and $\sum m_\nu = 3$ eV
from WMAP and SDSS data. From this figure it is obvious that
models with high neutrino mass are not ruled out by any single
data point, but rather by a general decrease in how well the
combined data fits. One fairly evident problem with the high
neutrino mass model is that the shape of the large scale structure
power spectrum becomes wrong. The model spectrum has too much
power at intermediate scales and too little at small scales.

In the upper part of this figure the deviation of the best fit
models from other cosmological data is shown. This data is not
used in deriving the best fit models, and therefore the figure
shows that the standard concordance model with $\sum m_\nu = 0$ is
not only a better fit to CMB and LSS data, but also more
consistent with other cosmological data.

\subsubsection{Combining measurements of $m_\nu$ and $N_\nu$.}

\begin{figure}[htbp]
\begin{center}
\epsfig{file=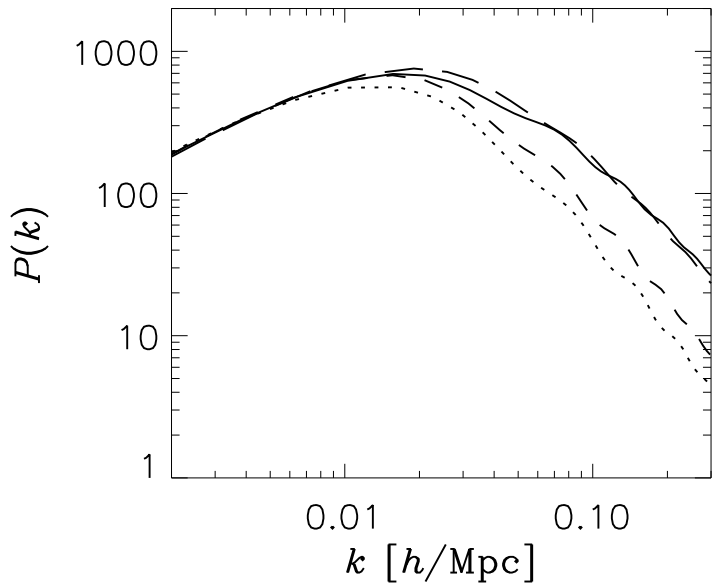,width=0.6\textwidth}
\end{center}
\bigskip
\caption{Power spectra for $\Lambda$CDM
  models with $\Omega_b = 0.05$, $\Omega = 1$, $h=0.7$, $n_s=1$, and
  $N_{\nu,{\rm massive}}=1$ and a common large-scale normalization.
  The full line is for $\Omega_\nu=0$, $\Omega_m=0.25$, $N_\nu=3$,
  dashed is for $\Omega_\nu=0.05$, $\Omega_m=0.25$, $N_\nu=3$, dotted
  is for $\Omega_\nu=0.05$, $\Omega_m=0.25$, $N_\nu=8$, and
  long-dashed is for $\Omega_\nu=0.05$, $\Omega_m=0.35$, $N_\nu=8$
  (from \protect\cite{hr04}).} \label{fig:pspec}
\end{figure}

The limits on neutrino masses discussed above apply only for
neutrinos within the standard model, i.e.\ three light neutrinos
with degenerate masses (if the sum is close to the upper bound).
However, if there are additional neutrino species sharing the
mass, or neutrinos have significant chemical potentials this bound
is changed. Models with massive neutrinos have suppressed power at
small scale, with suppression proportional to
$\Omega_\nu/\Omega_m$. Adding relativistic energy further
suppresses power at scales smaller than the horizon at
matter-radiation equality. For the {\it same} matter density such
a model would therefore be even more incompatible with data.
However, if the matter density is increased together with $m_\nu$,
and $N_\nu$, excellent fits to the data can be obtained. This
effect is shown in Fig.~\ref{fig:pspec}.

\begin{figure}[htbp]
\begin{center}
\epsfig{file=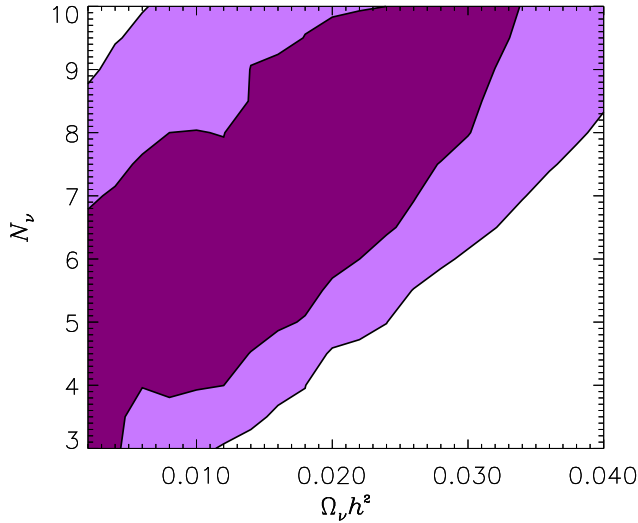,width=0.6\textwidth}
\end{center}
\bigskip
\caption{Likelihood contours (68\% and 95\%) for the case of
$N_\nu$ neutrinos with equal masses, calculated from WMAP and 2dF
data (from \protect\cite{hr04}).} \label{fig:mnunnu}
\end{figure}

The effect on likelihood contours for $(\Omega_\nu,N_\nu)$ can be
seen in Fig.~\ref{fig:mnunnu} which is for the case where $N_\nu$
species the total mass equally.

A thorough discussion of these models can be found in
Refs.~\cite{hr04,Crotty:2004gm}.

\subsubsection{Future neutrino mass measurements}

The present bound on the sum of neutrino masses is still much
larger than the mass difference, $\Delta m_{23} \sim 0.05$ eV
\cite{Fogli:2003th,Maltoni:2003da}, measured by atmospheric
neutrino observatories and K2K . This means that if the sum of
neutrino masses is anywhere close to saturating the bound then
neutrino masses must the almost degenerate. The question is
whether in the future it will be possible to measure masses which
are of the order $\Delta m_{23}$, i.e.\ whether it can determined
if neutrino masses are hierarchical.

By combining future CMB data from the Planck satellite with a
galaxy survey like the SDSS it has been estimated that neutrino
masses as low as about 0.1 eV can be detected
\cite{Hannestad:2002cn,pastor04}. Another possibility is to use
weak lensing of the CMB as a probe of neutrino mass. In this case
it seems likely that a sensitivity below 0.1 eV can also be
reached with CMB alone \cite{Kaplinghat:2003bh}.

As noted in Ref.~\cite{pastor04} the exact value of the
sensitivity at this level depends both on whether the hierarchy is
normal or inverted, and the exact value of the mass splittings.

\subsection{Neutrino warm dark matter}

While CDM is defined as consisting of non-interacting particles
which have essentially no free-streaming on any astronomically
relevant scale, and HDM is defined by consisting of particles
which become non-relativistic around matter radiation equality or
later, warm dark matter is an intermediate. One of the simplest
production mechanisms for warm dark matter is active-sterile
neutrino oscillations in the early universe
\cite{Hansen:2001zv,Abazajian:2001vt,Abazajian:2001nj,Shi:1998km,%
Dodelson:1993je}.

One possible benefit of warm dark matter is that it does have some
free-streaming so that structure formation is suppressed on very
small scales. This has been proposed as an explanation for the
apparent discrepancy between observations of galaxies and
numerical CDM structure formation simulations. In general
simulations produce galaxy halos which have very steep inner
density profiles $\rho \propto r^\alpha$, where $\alpha \sim
1-1.5$, and numerous subhalos
\cite{Kazantzidis:2003hb,Ghigna:1999sn}. Neither of these
characteristics are seen in observations and the explanation for
this discrepancy remains an open question. If dark matter is warm
instead of cold, with a free-streaming scale comparable to the
size of a typical galaxy subhalo then the amount of substructure
is suppressed, and possibly the central density profile is also
flattened
\cite{Yoshida:2003rm,Haiman:2001dg,Avila-Reese:2000hg,Bode:2000gq,%
Colin:2000dn,Hannestad:2000gt,jsd} . In both cases the mass of the
dark matter particle should be around 1 keV
\cite{Hogan:2000bv,Dalcanton:2000hn}, assuming that it is
thermally produced in the early universe.

On the other hand, from measurements of the Lyman-$\alpha$ forest
flux power spectrum it has been possible to reconstruct the matter
power spectrum on relatively small scales at high redshift. This
spectrum does not show any evidence for suppression at sub-galaxy
scales and has been used to put a lower bound on the mass of warm
dark matter particles of roughly 1.1 keV
\cite{Narayanan:2000tp,Colombi:1995ze}. An even more severe
problem lies in the fact that star formation occurs relatively
late in warm dark matter models because small scale structure is
suppressed. This may be in conflict with the low-$l$ CMB
temperature-polarization cross correlation measurement by WMAP
which indicates very early reionization and therefore also early
star formation. One recent investigation of this found warm dark
matter to be inconsistent with WMAP for masses as high as 10 keV
\cite{Yoshida:2003rm}.

The case for warm dark matter therefore seems quite marginal,
although at present it is not definitively ruled out by any
observations.

\subsection{Neutrinos as the source for high energy cosmic rays}

Nucleons with energy above the threshold for photo-pion production
on the CMB rapidly downscatter in energy, the mean free path being
of order 20-30 Mpc. At the present CMB temperature of 2.7 K the
threshold, known as the Greisen-Zatzepin-Kuzmin energy \cite{GZK},
is roughly $4 \times 10^{19}$ GeV. On the other hand a significant
number of particles with energies above the GZK energy have been
observed. At present there is some controversy about the number of
such particles observed between different experimental
collaborations using different techniques. The HiRes \cite{HiRes}
collaboration finds a decline in the number of events beyond the
GZK energy which is in fact compatible with a cut-off. On the
other hand the AGASA \cite{AGASA} collaboration finds that the
spectrum is consistent with no cut-off, and even with a hardening
of the spectrum at very high energies. Recent reviews of these
issues can be found in Ref.~\cite{sigl1}.

Either these particles must come from relatively nearby sources or
they are not nucleons (or nuclei) with standard model
interactions. One explanation which has been proposed is the
$Z$-burst scenario
\cite{Weiler:1982qy,Weiler:1997sh,ringwald1,ringwald2,Gelmini:2002xy}.
In this model, neutrino primaries with very high energy, $E \gg
E_{\rm GZK}$, annihilate with neutrinos in the cosmic background
to produce the observed protons. The cross section is
significantly enhanced at $E_{\rm CM} = m_Z$, corresponding to a
neutrino primary energy of $E \sim m_Z^2/2E_{\nu,0} \sim m_Z^2/2
m_\nu$. If neutrino masses are hierarchical then the largest mass
is of order $m_\nu \sim \Delta m_{\rm atm} \sim 0.05$ eV, meaning
that $E_\nu \sim 10^{23}$ eV. If neutrinos are produced by pion
decay in AGNs this must mean that particles of even higher
energies are produced there.

On the other hand, if neutrinos have masses close to saturating
the cosmological bound then the primary energy can be
significantly lower.

Another requirement is that the neutrino annihilation must take
place within the GZK sphere. This leads to a too low flux unless
the rate is somehow enhanced. It has previously been proposed that
this can be explained in models with significant neutrino chemical
potential \cite{gelmini}. However, the present cosmological bounds
on $\eta$ rule this out. If neutrinos are of eV mass then they do
have significant clustering on GZK scales which can also enhance
the rate by a factor of about 2. {\it If} ultra high energy cosmic
rays are explained by the $Z$-burst this means that a mass bound
on neutrinos can in principle be obtained. In
Refs.~\cite{ringwald1,ringwald2} it was estimated that if the
annihilations are within the galactic halo it requires a neutrino
mass of $m_\nu = 2.34^{+1.29}_{-0.84}$ eV, and if the annihilation
happens within the local supercluster the mass must be $m_\nu =
0.26^{+0.20}_{-0.14}$ eV. The first case is already ruled out, but
the second possibility might work. It has, however, also been
shown that the values obtained in \cite{ringwald1,ringwald2} are
strongly model dependent \cite{kalashev}.

At present the feasibility of the $Z$-burst scenario remains an
open question. One problem is that the annihilation process
produces a background of low energy gamma rays which may be in
conflict with EGRET observations, depending on the magnitude of
the local neutrino density enhancement.

In any case the $Z$-burst scenario is also interesting from the
possibility of getting an independent detection of the
cosmological neutrino background if the needed very high fluxes of
ultra-high energy neutrinos is measured in future detectors like
Auger.

%%%%%%%%%%%%%%%%%%%%%%%%%%%%%%%%%%%%%%%%%%%%%%%%%%%%%%%%%%%%%%%%%%%%%%
\section{Neutrinos in the very early universe - Leptogenesis} %%%%%%%%
%%%%%%%%%%%%%%%%%%%%%%%%%%%%%%%%%%%%%%%%%%%%%%%%%%%%%%%%%%%%%%%%%%%%%%

A particularly attractive model for baryogenesis involves massive,
right handed neutrinos, and is known as baryogenesis via
leptogenesis \cite{fukugita}. The basic idea is that the masses of
left handed Majorana neutrinos are generated from couplings to
very massive, right handed neutrinos.

These massive, right handed states are unstable and because they
are Majorana particles their decay violate lepton number.
Futhermore the decays are out of equilibrium and can violate CP
which means that all the Sakharov conditions for generating a net
lepton number are present. This lepton number can then
subsequently be transferred to the baryon sector via standard
model interactions and account for the observed baryon number of
the universe.

A particularly simple model for this is thermal leptogenesis where
the right handed neutrinos are equilibrated at high temperatures
directly via their interactions with the thermal plasma. In this
case, the correct baryon number is produced only if the following
conditions are fulfilled
\cite{Davidson,hamaguchi,buchmuller,buchmuller2,giudice,hambye}:
a) Masses of the light neutrinos must be less than about 0.1-0.15
eV, and b) Masses of the right handed neutrinos must be larger
than about $10^8$ GeV. The first condition is interesting because
it provides a strong, albeit very model dependent, bound on the
light neutrino masses. The second condition is interesting because
it is so high that it might be in conflict with the upper bound on
the reheating temperature in supersymmetric models. This bound
arises from overproduction and subsequent decay of gravitinos and
could probably be relaxed in models where the gravitino is the
lightest supersymmetric particle.

Taken at face value the thermal leptogenesis constraint on light
neutrino masses is the most restrictive cosmological limit known.
However, it is {\it not} a constraint at the same level as
experimental bounds, or even bounds from CMB and large scale
structure. The derivation involves a chain of assumptions: a)
Leptogenesis is the correct model of baryogenesis, b) leptogenesis
is thermal, c) The heavy neutrino masses are hierarchical. If
either of the first two assumptions are relaxed then there is
essentially no mass bound from this argument. If the last
assumption is relaxed then it has been shown in Ref.~\cite{hambye}
that the mass bound on the left handed neutrinos can be relaxed by
almost an order of magnitude.

With this in mind specific mass bounds on neutrino masses from
thermal leptogenesis should be taken as both interesting and
suggestive, but not as strict and generally applicable bounds.

%%%%%%%%%%%%%%%%%%%%%%%%%%%%%%%%%%%%%%%%%%%%%%%%%%%%%%%%%%%%%%%%%%%%%%
\section{Discussion} %%%%%%%%%%%%%%%%%%%%%%%%%%%%%%%%%%%%%%%%%%%%%%%%%
%%%%%%%%%%%%%%%%%%%%%%%%%%%%%%%%%%%%%%%%%%%%%%%%%%%%%%%%%%%%%%%%%%%%%%

In the present paper I have discussed how cosmological
observations can be used for probing fundamental properties of
neutrinos which are not easily accessible in lab experiments.
Particularly the measurement of absolute neutrino masses from CMB
and large scale structure data has received significant attention
over the past few years. In Table \ref{table:summary} I summarize
neutrino mass bounds from cosmological observations and other
astrophysical and experimental bounds.

\begin{table}
\caption{Summary table of cosmological neutrino mass limits. For
completeness bounds from other sources, astrophysical and
experimental, are also listed.}
\begin{center}
\begin{tabular}{@{}lll} \hline

Method & Bound on $\sum m_\nu$ & Data used \\
\hline

$\Omega_\nu h^2 \lesssim 0.15$  &   14 eV     &  $\Omega_\nu < \Omega_m$ \\

CMB and LSS     &   0.7-1 eV    &   WMAP, 2dF, SDSS \\

\hline

SN1987A         &   $m_{\nu_e} < 5-20$ eV  $(\bar{\nu}_e)$ &
SN1987A
cooling curve \cite{Kernan:1994kt,Loredo:2001rx}\\

$\beta$-decay   &   $m_{\nu_e} < 2.2$ eV $(\nu_e)$  & Mainz experiment \cite{kraus} \\

$0\nu 2\beta$-decay    & $m_{\nu,{\rm eff}} < 0.35$
eV & Heidelberg-Moscow \cite{Klapdor-Kleingrothaus:2000sn} \\

& 0.1 eV $< m_{\nu,{\rm eff}} <$ 0.9 eV & Heidelberg-Moscow
\cite{Klapdor-Kleingrothaus:2001ke,Klapdor-Kleingrothaus:2004wj}
\\

\hline
\end{tabular}
\end{center}
\label{table:summary}
\end{table}

Another cornerstone of neutrino cosmology is the measurement of
the total energy density in non-electromagnetically interacting
particles. For many years Big Bang nucleosynthesis was the only
probe of relativistic energy density, but with the advent of
precision CMB and LSS data it has been possible to complement the
BBN measurement. At present the cosmic neutrino background is seen
in both BBN, CMB and LSS data at high significance.

Finally, cosmology can also be used to probe the possibility of
neutrino warm dark matter, which could be produced by
active-sterile neutrino oscillations.

In the coming years the steady stream of new observational data
will continue, and the cosmological bounds on neutrino will
improve accordingly. For instance, it has been estimated that with
data from the upcoming Planck satellite it could be possible to
measure neutrino masses as low as 0.1 eV.

Certainly neutrino cosmology will continue to be a prospering
field of research for the foreseeable future.

%%%%%%%%%%%%%%%%%%%%%%%%%%%%%%%%%%%%%%%%%%%%%%%%%%%%%%%%%%%%%%%%%%%%%%
\section*{Acknowledgments} %%%%%%%%%%%%%%%%%%%%%%%%%%%%%%%%%%%%%%%%%%%
%%%%%%%%%%%%%%%%%%%%%%%%%%%%%%%%%%%%%%%%%%%%%%%%%%%%%%%%%%%%%%%%%%%%%%

I acknowledge use of the publicly available CMBFAST package
written by Uros Seljak and Matthias Zaldarriaga~\cite{CMBFAST} and
the use of computing resources at DCSC (Danish Center for
Scientific Computing). I also wish to thank Georg Raffelt for
valuable comments on the manuscript.

%%%%%%%%%%%%%%%%%%%%%%%%%%%%%%%%%%%%%%%%%%%%%%%%%%%%%%%%%%%%%%%%%%%%%%
\section*{References} %%%%%%%%%%%%%%%%%%%%%%%%%%%%%%%%%%%%%%%%%%%%%%%%
%%%%%%%%%%%%%%%%%%%%%%%%%%%%%%%%%%%%%%%%%%%%%%%%%%%%%%%%%%%%%%%%%%%%%%

\end{document}